\documentclass[twoside]{article}
\usepackage{fleqn,espcrc2}

\newcommand{\nn}{\nonumber\\}\newcommand{\p}[1]{(\ref{#1})}

\newcommand{\AmS}{{\protect\the\textfont2
  A\kern-.1667em\lower.5ex\hbox{M}\kern-.125emS}}

\title{The Space Filling Dirichlet 3-Brane in $N=2$, $D=4$
Superspace}

\author{I. Bandos
\address[KIPT]{Institute for Theoretical Physics, NSC KIPT\\
Akademicheskaya Street 1, 61108 Kharkov, Ukraine}
\address[B]{Dpto. de Fisica Teorica,
Facultad de Fisica,\\ Universidad de Valencia, E-46100-Burjassot
(Valencia), Spain},
P. Pasti\address[PD]{Universit\`a Degli Studi di Padova,
Dipartimento di Fisica ``Galileo Galilei''\\
ed INFN, Sezione Di Padova Via F. Marzolo, 8, 35131 Padova,
Italia},
 A. Pokotilov \address{
Physics Program, The Graduate Center, The City University of New
York, \\
365 Fifth Avenue, New York, NY 10016-4309 },
 D. Sorokin\addressmark[KIPT]\addressmark[PD]
                and
M. Tonin \addressmark[PD]
 }

\begin{document}

\begin{abstract}
We discuss a four--dimensional Volkov--Akulov supersymmetric
theory on a D3--brane with $N=2$ supersymmetry broken down to
$N=1$.
\vspace{1pc}
\end{abstract}

\maketitle

\section{INTRODUCTION}

In \cite{vak} Volkov and Akulov proposed the first
four--dimensional field theoretical model which possessed
space--time supersymmetry. The model was constructed in such a
way that (as we assume to be realized in nature) supersymmetry is
broken spontaneously with a ``neutrino" playing the role of the
associated fermionic Goldstone particle. This was the first
example of a mechanism of spontaneous breaking of global
supersymmetry which was generalized by Volkov and Soroka
\cite{vs} to the super--Higgs effect in the first supergravity
model with spontaneously broken local supersymmetry \cite{volkov}.

Later on it has been realized \cite{branes} that models of the
Volkov--Akulov type describe supersymmetric effective field
theories exhibiting partial supersymmetry breaking on the
worldvolumes of branes. This subject has recently faced a
significant revival of interest due to the extensive study of
various aspects of brane physics (see e.g.
\cite{bg,schw1,Kallosh:1997aw,breaking,bik,bik1}).

 The aim of this work is to study peculiarities of the
superembedding description \cite{stv,sem,GAP,hs,I,pr} of a space
filling D(irichlet)3--brane propagating in an $N=2$, $D=4$
superspace and to establish the relationship of this covariant
geometrical formulation with a Goldstone superfield formulation
of an $N=1$,
$D=4$ supersymmetric Dirac--Born--Infeld theory \cite{bg,bik,bik1} based on
methods of nonlinear realizations first applied to supersymmetry
by Volkov and Akulov.

We shall show that the superembedding conditions and worldvolume
gauge field constraints do not put D3--brane dynamics in $D=4$ on
the mass shell, in contrast, for example, to the case of a
D3--brane \cite{hs,bst} and a D9--brane
\cite{abkz} in type IIB D=10 supergravity.
A geometrical consequence of these conditions is the Grassmann
analyticity of both the $N=1$, $D=4$ superworldvolume and $N=2$,
$D=4$ target superspace. This is an extension to the D3--brane of
results on the relationship of the superembedding condition with
Grassmann analyticity properties of supermanifolds observed
previously in the cases of an $N=1$, $D=4$ superparticle and an
$N=1$, $D=4$ superstring \cite{stv,Ivanov:1991ub,Delduc:1992if,Delduc:1992fc}.

The fact that the superembedding conditions are off--shell
constraints will allow us to construct a worldvolume superfield
action for the space filling D3--brane which we briefly discuss
in the end of this contribution.

\medskip
\noindent
{\bf Notation}.
We use the formalism of two-dimensional Weyl spinors both in
$N=1$, $D=4$ superworldvolume and in $N=2$, $D=4$ target
superspace. Since the D3--brane in $D=4$ is a space filling brane
we can always gauge fix local Lorentz rotations in the
worldvolume in such a way that they coincide with Lorentz
transformations in the target superspace. So there is no need to
distinguish between the vector and spinor indices corresponding
to the tangent spaces of superworldvolume and target superspace.
Small letters of the Greek and Latin alphabet stand,
respectively  for spinor and vector indices, e.g.
$\alpha,
\dot\alpha=1,2; ~a, b=0,1,2,3$. Capital Latin letters denote both the
spinor and vector indices. The curved target superspace indices
denoted by the letters from the second half of the alphabets are
underlined to indicate that the worldvolume and target space
superdiffeomorphism groups are a priori independent.

\section{SUPEREMBEDDING CONDITIONS}

In the case of the space filling superbranes the basic
superembedding condition reads \cite{abkz} that the superembedding
of the brane superworldvolume into a target superspace is carried
out in such a way that (using local Lorentz transformations on the
worldvolume and in target superspace) it is always possible to
choose the vector component
$e^a$ of a worldvolume supervielbein
\begin{equation}\label{e}
e^A(z) =(e^a, e^\alpha, \bar e^{\dot\alpha}), \quad z^M=(\xi^m,
\eta^\mu,
\bar\eta^{\dot\mu})
\end{equation}
to coincide with the pullback of the vector component $E^a$ of a
target space supervielbein
\begin{eqnarray}\label{E}
E^{A}(Z)&=&(E^a, E^{I\alpha},
\bar E^{I\dot\alpha}), \quad \nn
Z^{\underline M}&=&(X^{\underline m},\Theta^{I\underline\mu},
\bar\Theta^{I\underline{\dot\mu}}), \quad I=1,2\,.
\end{eqnarray}
Namely,
\begin{equation}\label{vector}
e^a=E^a(Z(z)).
\end{equation}
Note that by imposing (\ref{vector}) we have identified the group
of local Lorentz rotations in the tangent space of the
superworldvolume with that of the target space Lorentz group,
while the worldvolume superdiffeomorphisms still remain an
independent group of transformations, and can be used (as we will
do in the final stage of our analysis), to impose a physical gauge
\begin{equation}\label{phys}
\xi^m=X^m, \quad \eta^\mu=\Theta^{1\mu}, \quad
\bar\eta^{\dot\mu}=
\bar\Theta^{1\dot\mu}.
\end{equation}
In this gauge the theory remains manifestly invariant under
$N=1$, $D=4$ supersymmetry associated with the
supertranslations along $\eta^\mu$ and
$\bar\eta^{\dot\mu}$, while the second target space
supersymmetry associated with the $\Theta^2$ translations is
realized nonlinearly in the transformation law of
$\Theta^{2\mu}(z)$, which implies its spontaneous breaking,
\begin{equation}\label{broken}
\delta
\Theta^{2}=\epsilon^2+i(\epsilon^2\sigma^a\bar\Theta^2+\Theta^2\sigma^a\bar\epsilon^2)
\partial_a\Theta^2.
\end{equation}
Thus $\Theta^{2\mu}(z)$ is the Volkov--Akulov Goldstone fermion
associated with the half of
$N=2$, $D=4$ supersymmetry spontaneously broken by the D3--brane.

As a consequence of (\ref{vector}) the pullback of $E^a$ along the
Grassmann directions (\ref{e}) of the superworldvolume is zero.
If the target superspace is flat (which is the case of our
interest)
\begin{equation}\label{Eflat}
E^{a}=dX^a-id\Theta^{I\alpha}\sigma^{a}_{\alpha\dot\alpha}\bar{\Theta}^{I\dot\alpha}+
i\Theta^{I\alpha}\sigma^{a}_{\alpha\dot\alpha}d\bar{\Theta}^{I\dot\alpha},
\end{equation}
and eq. (\ref{vector}) implies
$$
E^{~a}_{\alpha}={\cal D}_{\alpha}X^{a} -i{\cal
D}_{\alpha}\Theta^{I}\sigma^{a}\bar{\Theta}^{I}-
i\Theta^{I}\sigma^{a}{\cal D}_\alpha\bar{\Theta}^{I}=0,
$$
\begin{equation}\label{Ea}
E^{~a}_{\dot\alpha}={\bar{\cal D}}_{\dot\alpha}X^{a} -i{\cal
D}_{\alpha}\Theta^{I}\sigma^{m}\bar{\Theta}^{I}-
i\Theta^{I}\sigma^{a}{\bar {\cal
D}}_{\dot\alpha}\bar{\Theta}^{I}=0,
\end{equation}
where ${\cal D}_{\alpha}$ and ${\bar{\cal D}}_{\dot\alpha}$ are
worldvolume covariant derivatives.

As far as the spinor components of the worldvolume supervielbein
(\ref{e}) are concerned, by a simple redefinition they can always
be chosen to coincide with the pullback of one of the two spinor
components of the target space supervielbein (\ref{E}). For
instance, we can choose
\begin{equation}
\label{eal}
e^{\alpha}= E^{1\alpha}, \qquad
\bar{e}^{\dot{\alpha}}=
\bar{E}^{\dot{1\alpha}}.
\qquad
\end{equation}
(Such a choice reflects the possibility of gauge fixing
kappa--symmetry  by putting
$\Theta^1|_{\eta=0} =0$ in the conventional Green--Schwarz
formulation of the Dirichlet brane dynamics \cite{schw1}).

Then the generic expression for the pullbacks of the fermionic
one--forms
$E^{2\alpha}$ and $\bar E^{2\dot\alpha}$ in the worldvolume local frame
(\ref{vector}) and (\ref{eal}) is
 \begin{equation}
\label{gfsue}
E^{2\alpha}= E^{1\beta}h_\beta^{~\alpha}(z) +
\bar{E}^{1\dot{\alpha}} C_{\dot{\alpha} }^{~\alpha}(z)
+ E^a \psi_a^{\alpha}(z),
\end{equation}
$$
\bar{E}^{2
\dot{\alpha}}=
\bar{E}^{1\dot{\beta}}\bar{h}_{\dot{\beta} }^{~\dot{\alpha}}(z) +
{E}^{1{\alpha}} \bar{C}_{~{\alpha} }^{\dot{\alpha} }(z)
+ E^a \bar{\psi}_a^{\dot{\alpha} }(z).
$$

Note that in the flat target superspace
\begin{equation}\label{eflat}
E^{I\alpha}=d\Theta^{I\alpha}, \quad \bar E^{I\dot\alpha}=
d\bar\Theta^{I\dot\alpha},
\end{equation}
and
\begin{eqnarray}\label{flatcom}
h_\beta^{~\alpha}(z)&=&{\cal D}_\beta\Theta^{2\alpha}, \quad
\psi_a^{\alpha}(z)={\cal D}_a\Theta^{2\alpha},\nn
C_{\dot{\alpha} }^{~\alpha}(z)&=&\bar{\cal
D}_{\dot\alpha}\Theta^{2\alpha},
\end{eqnarray}
the analogous expressions being valid for the complex conjugate
superfields $\bar h$, $\bar C$ and $\bar \psi$.

The possibility of identifying  the worldvolume supervielbein
with the ``pulled back" components (\ref{vector}) and (\ref{eal})
of the target space supervielbein implies that in flat target
superspace the induced superworldvolume geometry is also flat,
and that the worldvolume spin connection is zero. This is
natural, since the brane worldvolume completely fills in (or
coincides with) the bosonic core of the flat target superspace. We
should note that though the superworldvolume is flat the
supervielbein
$e^A$ defined by (\ref{vector}) and (\ref{eal}) differs from the
standard flat superspace basis
\begin{equation}\label{e0}
e^A_0=(d\xi^a-id\eta\sigma^a\bar\eta+i\eta\sigma^ad\bar\eta,
~d\eta^\alpha, ~d\bar\eta^{\dot\alpha}).
\end{equation}
This, in particular, implies that the superworldvolume covariant
derivatives
${\cal D}_A$ in (\ref{Ea}) and (\ref{flatcom}) associated with the basis
(\ref{vector}) and (\ref{eal}) differ from conventional flat
covariant derivatives and form a more complicated superalgebra,
which we shall present a bit later.

The integrability of (\ref{gfsue}) and (\ref{eflat}) requires
some differential relations between the components $h(z),~ C(z)$
and $\psi(z)$ of the superforms (\ref{gfsue}), however the
integrability of the superembedding condition (\ref{vector}),
which implies that the worldvolume torsion is the pullback of the
target space torsion
\begin{eqnarray}\label{Ta}
de^a=dE^a\equiv T^a&&\nn
=- 2 i E^{\alpha 1}
\wedge
\bar{E}^{\dot{\alpha} 1}
\sigma^a_{\alpha\dot{\alpha}}&
- & 2 i E^{\alpha 2}
\wedge
\bar{E}^{\dot{\alpha} 2}
\sigma^a_{\alpha\dot{\alpha}},
\end{eqnarray}
does not put any further restrictions on $h$, $C$, and $\psi$,
and these superfields are still too general to be associated with
the physical modes of the D3--brane, which form a gauge vector
supermultiplet.

The situation when the basic superembedding condition is not
enough to determine the dynamics of the brane even off the mass
shell is generic for the space--filling \cite{abkz} and
codimension one
\cite{Howe:2000vk} branes. In such cases, for the superembedding to describe
superbrane dynamics, an additional constraint should be imposed
\footnote{Note that these additional constraints are reproduced
by the generalized action \cite{GAP,bst} on the same footing as
the basic superembedding conditions and the dynamical equations of
motion.}. In our case this is a constraint on an `extended'
field--strength two--form
\begin{equation}\label{F}
F_{2}=dA-B_2
\end{equation}
of a worldvolume gauge field $dz^MA_M(z)$ living on the D3-brane,
 the two--form $B_{2}$ being the pullback of an
``NS--NS'' gauge superfield of an $N=2$, $D=4$ supergravity which
the D3 brane couples to.

\subsection{Worldvolume gauge field constraints}
We assume the worldvolume superfield constraint on $F_{2}$ to be
\begin{equation}
\label{gc}
F_{2}=dA-B_2 = {1\over 2} E^a \wedge E^b  F_{ba},
\end{equation}
which implies that $F_2$ has nonzero components only along bosonic
directions (\ref{vector}) of the superworldvolume.

To argue that the constraint (\ref{gc}) is relevant to the
description of the Born--Infeld gauge field and of the D3--brane
as a whole, we note that it follows from a generalized action
\cite{GAP} for super--Dp--branes constructed and analyzed in
\cite{bst,abkz}, and in a linear approximation it reduces to
standard $N=1$, $D=4$ super--Maxwell constraints, as we shall
demonstrate in Subsection 2.3 upon analyzing the integrability
condition of (\ref{gc})
 \begin{equation}
\label{intgc}
-H_3 =  E^a \wedge T^b F_{ba} +
{1\over 2} E^a \wedge
E^b \wedge dF_{ba},
\end{equation}
where $H_3=dB_2$ and $T^b$ are the pullbacks of, respectively,
the $B_2$ field strength and the torsion of the target superspace
which we will further consider to be flat (see eqs.
(\ref{Eflat}), (\ref{eflat}) and (\ref{Ta})). In flat target
superspace $H_3$ has the following form
\begin{equation}
\label{H3}
H_3 = 2 i E^{a}
\wedge
(E^{1\alpha}
\wedge
\bar{E}^{1 \dot{\alpha}}
- E^{2\alpha}
\wedge
\bar{E}^{2\dot{\alpha}})
\sigma_{a\alpha\dot{\alpha}},
\qquad
\end{equation}
and from \p{intgc} and \p{Ta} we get
\begin{eqnarray}\label{intgc1}
&& - 2 i E^{a}
\wedge
(E^{\alpha 1}
\wedge
\bar{E}^{\dot{\alpha} 1}
\sigma^b_{\alpha\dot\alpha} (\eta -F)_{ba} \nn
&&- E^{\alpha 2}
\wedge
\bar{E}^{\dot{\alpha} 2} \sigma^b_{\alpha\dot\alpha}
(\eta +F)_{ba} ) = \nn
&&{1\over 2} E^a \wedge E^b \wedge dF_{ba}.
\end{eqnarray}

Our goal is to show that when the constraints ({\ref{vector}) and
(\ref{gc}) are imposed the components of the superfields
$F_2(z)$, $h(z)$, $C(z)$ and $\psi(z)$ either vanish or are
expressed through the worldvolume chiral spinor superfield
$\Theta^2_\alpha(z)$ describing the gauge vector supermultiplet.
To this end we analyze the integrability condition (\ref{intgc1}).

Substituting \p{gfsue} into
\p{intgc1} and taking its $E^{\alpha 1}\wedge E^{\beta 1}$ and
$\bar E^{\dot\alpha 1}\wedge \bar E^{\dot\beta 1}$
components one finds that
 \begin{eqnarray}
\label{Ea2}
h_{(\alpha}^{~\gamma}
\sigma^b_{\gamma\dot{\gamma}}
\bar{C}_{~\beta )}^{\dot{\gamma}}
(\eta -F)_{ba} =0, &\nn
\bar h_{(\dot\alpha}^{~\dot\gamma}
\sigma^b_{\dot\gamma{\gamma}}
{C}_{\dot\beta)}^{~{\gamma}} (\eta -F)_{ba} =0,&
\end{eqnarray}
where $\eta_{ab}$ is the $D=4$ Minkowski metric.

If the  matrix
$(\eta -F)$ is non--degenerate (which is the general assumption
of the Born--Infeld--like models) the equations (\ref{Ea2}) are
satisfied if and only if
\begin{equation}
\label{sol1}
\bar{C}_{~\beta}^{\dot{\gamma}} =0, \qquad
{C}_{\dot\beta}^{~{\gamma}}=0
\end{equation}
or
\begin{equation}
\label{sol2}
h_\alpha^{~\beta} = 0, \qquad \bar h_{\dot\alpha}^{~\dot\gamma}=0.
\end{equation}
As one can verify the second choice (eq. (\ref{sol2})) leads to a
trivial solution of the superembedding conditions (which does not
describe any physical dynamical system), so we shall analyze the
nontrivial consequences of the first solution (\ref{sol1}). Then
the spinor supervielbein pullbacks (\ref{gfsue}) take the form
 \begin{eqnarray}
\label{gfsue1}
E^{2\alpha}= E^{1\beta}h_\beta^{~\alpha} + E^a
\psi_a^{\alpha},&&\nn
\qquad
\bar{E}^{2
\dot{\alpha}}=
\bar{E}^{1\dot{\beta}}\bar{h}_{\dot{\beta} }^{~\dot{\alpha} }
+ E^a \bar{\psi}_a^{\dot{\alpha} }.&&
\end{eqnarray}

Now consider the $E^{1\alpha}\wedge \bar E^{1\dot\alpha}$
component of (\ref{intgc1}). In view of (\ref{gfsue1}) it reduces
to
\begin{equation}
\label{hshk}
h_\alpha^{~\beta}
\sigma^{a}_{\beta\dot{\beta}}
\bar{h}_{\dot{\alpha} }^{~\dot{\beta} } \equiv
(h\sigma^a \bar{h})_{\alpha\dot{\alpha}}=
\sigma^{b}_{\alpha\dot{\alpha}} k_{b}^{~a},
\end{equation}
where the matrix $k_{b}^{~a}$ takes values in the
(pseudo)orthogonal group $SO(1,3)$ which follows from its
definition
$$
k_{b}^{~a} = (\eta - F)_{bc}(\eta + F)^{-1~ca}
$$
$$
 = (\eta +
F)^{-1}_{bc}(\eta - F)^{ca}= \delta_{b}^{~a}-2((\eta +
F)^{-1}F)_{b}^{~a},
$$
\begin{equation} \label{kin} k^{-1}=k^T, \quad
(\delta_b^{~a}+k_b^{~a})= 2 \eta_{bc}(\eta+F)^{-1}{}^{ca}.
\end{equation}
Then the relation (\ref{hshk}) implies that (up to a
$U(1)$ rotation)
$h_\alpha^{~\beta}$ belongs
to a spinor representation of $SO(1,3)$
$$
h_\alpha^{~\beta}
\qquad \in \qquad SL(2,C) \times U(1)
$$
and
\begin{equation}\label{hin}
| det(h)| = 1 ~\rightarrow ~det(h) = e^{2 i a(z) },
\end{equation}
where the real superfield $a(z)$ takes values on the  circle
$S^1$.

We have thus found the relationship between the spin--tensor
superfield $h_\alpha^{~\beta}(z)$ (which in the flat target
superspace is
$h_\alpha^{~\beta}(z)={\cal D}_\beta\Theta^{2\alpha}$)
and the field strength $F_{ab}(z)$ of the worldvolume gauge
field. Namely, from (\ref{hshk}) and (\ref{kin}) it follows that
\begin{equation}\label{Fa}
F_{a}^{~b}=H^{-1~b}_{~a}-\delta_a^{~b},
\end{equation}
{\rm where}
 $H_a^{~b}={1\over
 2}[{1\over 2}tr (h\sigma_a
 \bar h\sigma^b)+\delta_a^{~b}].$

The equation (\ref{hin}) implies that the superfield
$h_{\alpha}^{~\beta}$ satisfies the nonlinear constraint
\begin{equation}\label{nonli}
det(h_\alpha^{~\beta})\cdot det(\bar
h_{\dot\alpha}^{~\dot\beta})=1.
\end{equation}

This is the exact form of the nonlinear generalization of the
Maxwell superfield constraint (see eq. (\ref{scalar}) below)
which was found  in \cite{bg} to order ${\cal O}(\Theta^3)$.

Because of the group theoretical properties (\ref{hshk}),
(\ref{kin}) and (\ref{hin})  of $h_\alpha^{~\beta}$ and
$k_b^{~a}$ they also satisfy the following relation
\begin{equation}\label{hdh}
h^{-1}{}^{~\gamma}_\alpha d h_\gamma^{~\beta}={1\over
2}(k^{-1}dk)^{ab}\sigma_{ab\alpha}^{~~~\beta}+ida(z)\delta^{~\beta}_\alpha.
\end{equation}
Eq. \p{hdh} implies, in particular,
\begin{equation}\label{tr(hdh)}
h^{-1}{}^{~\gamma}_\beta d h_\gamma^{~\beta}= 2ida(z).
\end{equation}

To conclude the analysis of the consequences of the superembedding
condition (\ref{vector}) and of the gauge field constraint
(\ref{gc}) we shall now demonstrate that they do not put the
theory on the mass shell.

The dynamical fermionic equation of motion of the D3--brane which
is obtained by varying the Green--Schwarz--like \cite{dbranes} or
the generalized action \cite{bst} for the D3--brane with respect
to
$\Theta^2$ is
\begin{eqnarray}\label{bDalh}
\sigma^b_{\alpha{\dot{\alpha}}}
{(\eta - F)^{-1}}_b^{~a} {\cal D}_a\Theta^{2\alpha}&\nn
\equiv
\sigma^b_{\alpha{\dot{\alpha}}}
{(\eta - F)^{-1}}_b^{~a} \psi_a^{\alpha} =0, &
\end{eqnarray}
(plus its complex conjugate).

We should, therefore, check that eq. (\ref{bDalh}) does not follow
from the constraints (\ref{vector}) and (\ref{gc}). To this end
let us note that in view of (\ref{vector}), (\ref{eal}),
(\ref{Ta}), (\ref{sol1}), (\ref{gfsue1}) and (\ref{hshk}) the
algebra of the worldvolume covariant derivatives ${\cal D}_A$ is
\begin{eqnarray}\label{susy1}
\{{\cal D}_\alpha,\bar{\cal
D}_{\dot\alpha}\}=-T^a_{\alpha\dot\alpha}{\cal D}_a=
2i\sigma^b_{\alpha\dot\alpha}(\delta_b^{~a}+k_b^{~a}){\cal
D}_a&\nn
 =4i \sigma_{b\alpha\dot\alpha}(\eta+F)^{-1}{}^{ba}{\cal
D}_a,&
\end{eqnarray}
\begin{equation}\label{susy2}
\{{\cal D}_\alpha,{\cal
D}_{\beta}\}=0=\{\bar{\cal D}_{\dot\alpha},\bar{\cal
D}_{\dot\beta}\},
\end{equation}
\begin{equation}\label{susy3}
\{{\cal D}_\alpha,{\cal
D}_{b}\}=-T^a_{\alpha b}{\cal D}_a =-2i (h\sigma^a
)_{\alpha\dot\alpha }
\bar{\psi}_b^{\dot\alpha}{\cal D}_a, \quad
\end{equation}
$$
\{\bar{\cal D}_{\dot\alpha},{\cal
D}_{b}\}=-T^a_{\dot\alpha b}{\cal D}_a =-2i
(\sigma^a\bar{h})_{\alpha\dot\alpha }
{\psi}_b^{\alpha}{\cal D}_a,
$$
\begin{equation}\label{susy4}
\{{\cal D}_b,{\cal
D}_{c}\}=-T^a_{bc}{\cal D}_a=-4i (\psi\sigma^a\bar{\psi}){\cal
D}_a.
\end{equation}

Then applying $\bar{\cal D}_{\dot\beta}$ to
$h_\alpha^{~\beta}$ of (\ref{flatcom}), and taking into account
(\ref{sol1}) and (\ref{susy1}) we find that
\begin{equation}
\label{bDalh1}
\bar{\cal D}_{\dot{\beta}}
h_{\beta}^{~\alpha} = 4i \sigma^b_{\beta{\dot{\beta}}} (\eta +
F)^{-1}_{ba}
\psi^{a\alpha},
\end{equation}
which relates $\psi^{a\alpha}$ with $h_{\beta}^{~\alpha}$ and
$F_{ab}$.

Now multiplying eq. (\ref{bDalh1}) by $(h^{-1})^{~\beta}_\alpha
(\bar h^{-1})^{~\dot\beta}_{\dot\alpha}$, and using the relations
(\ref{hshk}), (\ref{kin}) and (\ref{tr(hdh)}) we get
 \begin{equation}\label{trI12'}
(\sigma_b)_{\alpha\dot\alpha}(\eta-F)^{-1}{}^{ba}
\psi_a{}^\alpha = -{1\over 4}i \bar{h}^{-1}{}_{\dot{\alpha}}^{~\dot{\beta}}
\bar{\cal D}_{\dot\beta}a(z),
\end{equation}
where the left hand side is the same as in eq. (\ref{bDalh}), but
it is non--zero, since $a(z)$ is generically non--constant. Thus,
in the case of the space--filling D3--brane the superembedding
conditions and the field strength constraint does not produce
dynamical equations of motion and, therefore, leave the theory off
the mass shell. The equations of motion arise only if in addition
we put $da(z)=0$ or $a(z)=const$. Then, on the mass shell, the
spin tensor $h$ becomes  an $SL(2,C)$ valued matrix (c.f.
\cite{abkz} for a D=10 super-D9-brane)
\begin{equation}\label{deth}
\det{h_{\alpha}^{~\beta}}=1.
\end{equation}
Eq. (\ref{deth}) can be regarded as the nonlinear superfield
equation of motion of the D3-brane, which generalzes the linear
super-Maxwell equation of motion (see Subsection 2.3).

\subsection{Grassmann analyticity}
As we have already mentioned, the superembedding conditions
({\ref{vector}), (\ref{Eflat}), (\ref{Ea}),
(\ref{eal})--(\ref{flatcom}) and the gauge field constraint
(\ref{gc}) result in double analyticity, i.e. Grassmann
analyticity both in the worldvolume and in target superspace, the
phenomenon which was declared in \cite{Ivanov:1991ub} as a
principle for some types of superembeddings, describing for
instance certain superparticles
 and superstrings \cite{stv,Ivanov:1991ub,Delduc:1992if,Delduc:1992fc}.
Indeed, since the integrability of the constraints requires eq.
(\ref{sol1}), from (\ref{eal})--(\ref{flatcom}) it follows that
$\Theta^{I\alpha}$ are chiral worldvolume superfields, i.e.
\begin{equation}\label{chiral}
{\bar{\cal D}}_{\dot\alpha}\Theta^{I\alpha}=0, \qquad
{\cal D}_{\alpha}\bar\Theta^{I\dot\alpha}=0.
\end{equation}
Then eqs. (\ref{Ea}) take the form
\begin{eqnarray}\label{Ea1}
{\cal D}_{\alpha}(X^{a} -i\Theta^{I}\sigma^{a}\bar{\Theta}^{I})=0,&\nn
{\bar{\cal D}}_{\dot\alpha}(X^{a}
+ i\Theta^{I}\sigma^{a}\bar{\Theta}^{I})=0,&
\end{eqnarray}
or
\begin{eqnarray}\label{chiralX}
X^a&=&{1\over 2}(X_R^a+X_L^a),\nn
X_L^a&-&X_R^a-2i\Theta^{I}\sigma^{a}\bar{\Theta}^{I}=0,
\end{eqnarray}
where $X^a_R=X^{a} -
i\Theta^{I}\sigma^{a}\bar{\Theta}^{I}=\overline{(X^a_L)}$ are
complex conjugate chiral worldvolume superfields
\begin{equation}\label{chiralX1}
{\bar{\cal D}}_{\dot\alpha}X^a_L=0, \qquad
{\cal D}_{\alpha}X^{a}_R=0.
\end{equation}

The equations (\ref{chiralX}) are nothing but the definition of
complex coordinates
$Z^M_L=(X_L^a=X^a+i\Theta^{I}\sigma^{a}\bar{\Theta}^{I}, \Theta^{I\alpha})$ of a chiral
subspace of the $N=2$, $D=4$ superspace, which in their turn are
chiral superfields in the $N=1$, $D=4$ superworldvolume.

We have thus obtained that the conditions imposed on the
embedding of the D3--brane imply that the superembedding is
performed in such a way that the chiral subsuperspace of the
superworldvolume gets mapped into the chiral subsuperspace of the
target superspace.

\subsection{Linearized limit}
We shall now demonstrate that in the physical gauge (\ref{phys})
and in a linearized limit in worldvolume superfields the gauge
field constraint (\ref{gc}) gives rise to standard constraints on
the field strength of the Maxwell field supermultiplet.

Upon imposing the physical gauge (\ref{phys}) the only
independent (chiral) variable which remains in the model is the
Volkov--Akulov Goldstone superfield
$\Theta^{2\alpha}(z)$, to which the gauge field strength $F_{ab}(z)$ is
related via eq. (\ref{Fa}).

To be able to perform a correct linearization limit we should
choose $\Theta^{2\alpha}(z)$ in the form
\begin{equation}\label{split}
\Theta^{2\alpha}(z)=\eta^\alpha+W^\alpha(z),
\end{equation}
where $W^\alpha(z)$ is a chiral worldvolume superfield.

This choice can be understood with the following reasoning. When
there is no a gauge field on the D3--brane worldvolume
$F_{ab}(z)=0$. Then the integrability (\ref{intgc1}) of the gauge
field constraint (\ref{gc}) reduces to
\begin{equation}\label{F=0}
2 i E^{a}\wedge (E^{1\alpha}
\wedge
\bar{E}^{1 \dot{\alpha}}
- E^{2\alpha}
\wedge
\bar{E}^{2\dot{\alpha}})
\sigma_{a\alpha\dot{\alpha}}=0,
\qquad
\end{equation}
which is satisfied if we choose
$E^{1\alpha}=E^{2\alpha}$ along superworldvolume.
Hence, in the static gauge this ``vacuum'' configuration of the
D3--brane can be associated with the map
$\eta^\alpha=\Theta^{1\alpha}=\Theta^{2\alpha}$, and
fluctuations around this solution are described by the chiral
superfield $W^\alpha(z)$ of eq. (\ref{split}).

We shall now analyze, in the static gauge (\ref{phys}), the
consequences of the superembedding (chirality) conditions
(\ref{chiral}), (\ref{Ea1}) and the integrability condition
(\ref{intgc1}) in the linear order in the fields
$W^\alpha(z)$ and $F_{ab}(z)$. From (\ref{phys}), (\ref{chiral}),
(\ref{Ea1}) and (\ref{split}) we find that in the linear
approximation
\begin{eqnarray}\label{calDal}
{\cal D}_\alpha=D_\alpha+ i(\sigma^a\bar W)_\alpha\partial_a+
i D_\alpha W\sigma^a\bar\eta\partial_a,&\nn
{\bar{\cal D}}_{\dot\alpha}=\bar D_{\dot\alpha}+
i(W\sigma^a)_{\dot\alpha}\partial_a + i \eta\sigma^a\bar
D_{\dot\alpha}\bar W\partial_a,&
\end{eqnarray}
where
\begin{eqnarray}\label{flatDal}
D_\alpha={\partial\over{\partial\eta^\alpha}}
+2i(\sigma^a\bar\eta)_\alpha{\partial\over{\partial\xi^a}},&
\nn
\bar D_{\dot\alpha}={\partial\over{\partial\bar\eta^{\dot\alpha}}}
+2i(\eta\sigma^a)_{\dot\alpha}{\partial\over{\partial\xi^a}}&
\end{eqnarray}
are flat covariant derivatives.

Note that in the linear approximation $W^\alpha(z)$ satisfies the
flat superspace chirality condition
\begin{equation}\label{flatchir}
\bar D_{\dot\alpha}W^{\alpha}=0, \qquad D_\alpha\bar
W^{\dot\alpha}=0.
\end{equation}

Finally eq. (\ref{Fa}) reduces to
\begin{equation}\label{linhsh}
F_{ab}={1\over
4}\sigma_{b}^{\alpha\dot\alpha}(\sigma_{a\beta\dot\alpha}D_{\alpha}W^{\beta}+
\sigma_{a\alpha\dot\beta}\bar{D}_{\dot\alpha}\bar{W}^{\dot\beta}),
\end{equation}
which, in particular, implies that
\begin{equation}\label{scalar}
D^{\alpha}W_{\alpha}+\bar{D}_{\dot\alpha}\bar{W}^{\dot\alpha}=0.
\end{equation}
In equations (\ref{flatchir}) and (\ref{scalar}) one can
recognize the standard constraints on the field strength
superfield of a Maxwell supermultiplet. They arise as the linear
approximation of the Goldstone superfield constraints
(\ref{chiral}) and (\ref{nonli}).

The Maxwell superfield equations of motion
\begin{equation}\label{Meq}
D^{\alpha}W_{\alpha}-\bar{D}_{\dot\alpha}\bar{W}^{\dot\alpha}=0.
\end{equation}
arise as the linearized approximation of the D3-brane superfield
equation (\ref{deth}).

We have thus demonstrated that the choice of the basic
superembedding condition (\ref{vector}) and the gauge field
constraint (\ref{gc}) is consistent with the linearized limit of
the D3--brane model which is  $N=1$, $D=4$ supersymmetric Maxwell
theory.

\section{The D3--brane action}
We now present a worldvolume superfield action which we assume to
produce upon integrating over Grassmann coordinates and solving
for the auxiliary fields the standard action \cite{dbranes} for
the D3--brane coupled to an $N=2$ supergravity.

The D3--brane couples to supergravity fields via the worldvolume
pullback of the Wess--Zumino form \cite{dbranes}
\begin{equation}\label{WZ}
\hat C=C_4+F_2\wedge C_2+{1\over 2}F_2\wedge F_2C_0,
\end{equation}
where $F_2$ is defined in (\ref{F}) and $C_p$ (p=0,2,4) are
`Ramond-Ramond' p--form fields.

Since, as we have shown in Subsection 2.2, the superembedding
conditions imply chirality of the worldvolume superfields we
assume the action to be an integral over $N=1$, $D=4$ chiral
superspace $Z_L=(\xi^m_L, \eta^\alpha)$ of an appropriate
pullback component of $\hat C$. Such a structure is prompted by
the form of the worldvolume superfield actions for a heterotic
string \cite{hstring} and a supermembrane
\cite{membra}. Because of the dimensional
reasons the Lagrangian is constructed with the use of
$\hat C_{\dot\alpha\dot\beta ab}$, and the action
(accompanied by the superembedding condition (\ref{gc})) has the
following formally simple form
\begin{equation}\label{action}
S=\int d^2\xi_L d^2\eta {\cal E}_L \sigma^{ab\dot\alpha\dot\beta}
\hat C_{\dot\alpha\dot\beta ab}~~+~~h.c.
\end{equation}
where ${\cal
E}_L=sdet(e^{~A}_M)~det(\eta_{ab}-F_{ab})~det^{-1}h^{~\alpha}_\beta$
is the chiral measure ${\cal D}_{\dot\alpha}
{\cal E}_L=0$, and $\sigma^{ab}$
is the antisymmetrized product of the Pauli matrices.

 Upon integration over $\eta$ (\ref{action}) should produce both the
Dirac--Born--Infeld and the Wess--Zumino term of the standard
D3--brane action. In the static gauge (\ref{phys}) and in the
linearized limit (\ref{split}), (\ref{flatchir}) and
(\ref{scalar}) the action (\ref{action}) reduces to the
superfield Maxwell action.

\section{CONCLUSION}
Using the superembedding approach we have shown that the
off--shell dynamics of the D3--brane in $N=2$, $D=4$ target
superspace is described by the worldvolume superfield
(superembedding) conditions (\ref{vector}) and (\ref{gc}).  In
the static gauge they reduce to nonlinear off--shell constraints
on the spinor (Goldstone) superfield strength of the
Dirac--Born--Infeld supermultiplet which generalize the Maxwell
superfield linear constraints. This establishes the link of the
superembedding formulation of the D3--brane with the nonlinear
realization method used by Bagger and Galperin \cite{bg} to
construct the $N=1$, $D=4$ superfield formulation of the
Dirac--Born--Infeld theory as a Volkov--Akulov--type model
exhibiting partial breaking of $N=2$ supersymmetry down to $N=1$.

The detailed analysis of the D3--brane superworldvolume action
(\ref{action}) and its relation to the Goldstone--Maxwell
superfield action and equations of motion of \cite{bg,bik} will be
given elsewhere.

\section*{Acknowledgements}
The authors are grateful to E. Ivanov and S.~Krivonos for
interest to this work and helpful discussions. This work was
partially supported by the INTAS, the European Commission RTN
Programme HPRN-CT-2000-00131 to which  P.P., D.S. and M.T. are
associated, and by the Ministerio de Educaci\'on y Cultura de
Espa$\tilde{n}$a (I.B.). I.B. and D.S. also acknowledge partial
support from the Grant N 2.51.1/52-F5/1795-98 of the Ukrainian
Ministry of Science and Technology.


\begin{thebibliography}{99}
\bibitem{vak}
D. V. Volkov and V. P. Akulov, JETP Letters 16 (1972) 438; Phys.
Lett. B46 (1973) 109.
\bibitem{vs}
D. V. Volkov and V. A. Soroka, JETP Letters 18 (1973) 312.
\bibitem{volkov}
D. V. Volkov, Supergravity before 1976, in "History of Original
Ideas and Basis Discoveries in Particle Physics", H. B. Newman
nad T. Ypsilantis (eds.), NATO ASI Series B Physics Vol. 352 pp.
663-674, Plenum Press, New York, 1996.
\bibitem{branes}
J. Hughes, J. Liu and J. Polchinski,  Phys. Lett.   B180 (1986) 370;\\
J. Hughes and J. Polchinski,  Nucl. Phys.  B278 (1986) 147.\\
A. Achucarro, J. Gauntlett, K. Itoh and P.K. Townsend,
 Nucl. Phys. B314 (1989) 129;\\
J. P. Gauntlett, K. Itoh and P. K. Townsend, Phys. Lett. B238
(1990) 65.
\bibitem{bg}
J. Bagger and A. Galperin,  Phys. Rev.  D55 (1997) 1091.
\bibitem{schw1}
M. Aganagic, C. Popescu and J.H. Schwarz,
Nucl.Phys.B490 (1997) 202.
\bibitem{Kallosh:1997aw}
R.~Kallosh, Volkov-Akulov theory and D-branes,hep-th/9705118. D.
V. Volkov Memorial Volume, V. Akulov and J. Wess (eds.) Lecture
notes in physics Vol. 509, p. 49. Springer--Verlag, Berlin,
Heidelberg, 1998.
\bibitem{breaking}
F. Gonzalez--Rey, I. Y. Park and M. Ro\~cek,  Nucl. Phys. B544
(1999) 243;\\
M. Ro\~cek and A. Tseytlin,  Phys. Rev.  D59 (1999)
106001.\\
 S. Bellucci, E. Ivanov and S. Krivonos, Fortsch.
Phys. 48 (2000) 19;  Phys. Lett.
B460 (1999) 348;\\
E. Ivanov and S. Krivonos,  Phys. Lett.
 B453 (1999)
237.\\
S. V. Ketov, Mod. Phys. Lett. A14 (1999) 501;
 Nucl. Phys. B553 (1999) 250; ``N=2 Super--Born--Infeld
Theory Revisited'', hep--th/0005126.\\
F. Delduc, E. Ivanov and S. Krivonos,  Nucl. Phys. B576
(2000) 196. \\
P. West, ``Automorphisms, Non-linear Realizations and
Branes'',  hep-th/0001216.\\
N.~D.~Lambert and P.~C.~West, ``Goldstone soliton interactions and
brane world neutrinos'', hep-th/0012121.
\bibitem{bik}
S. Bellucci, E. Ivanov and S. Krivonos, Phys. Lett. B482 (2000)
233.
\bibitem{bik1}
S. Bellucci, E. Ivanov and S. Krivonos, ``Superbranes and Super
Born-Infeld Theories from Nonlinear Realizations". In these
proceedings. hep-th/0103136.
\bibitem{stv}
D. Sorokin, V. Tkach and D. V. Volkov, Mod. Phys. Lett.
 A4 (1989) 901;\\
D. Sorokin, V. Tkach, D. V. Volkov and A. Zheltukhin, Phys. Lett.
B216 (1989) 302.
\bibitem{sem}
I.\  Bandos, P.\  Pasti, D.\  Sorokin, M.\ Tonin and D.\ Volkov,
Nucl.Phys.  B446 (1995) 79.
\bibitem{GAP}
I.\ Bandos, D.\ Sorokin and D.\ Volkov, Phys. Lett. B352 (1995)
269.
\bibitem{hs}
P.S. Howe  and E. Sezgin,  Phys.Lett. B390 (1997) 133; Ibid. B394
(1997) 62.
\bibitem{I}
I. Bandos,
{\em Lectures Notes in Physics} 524 (1999) 146, hep-th/9807202.
\bibitem{pr}
D. Sorokin, Physics Reports 329 (2000) 1.
\bibitem{bst}
I. Bandos, D. Sorokin and M. Tonin,
 Nucl.Phys. B497 (1997) 275.
\bibitem{abkz}
V. Akulov, I. Bandos, W.Kummer and V. Zima, Nucl. Phys.
 B527 (1998) 61.
\bibitem{Ivanov:1991ub}
E.~A.~Ivanov and A.~A.~Kapustnikov,
Phys.\ Lett.\ B {\bf 267} (1991) 175.
\bibitem{Delduc:1992if}
F.~Delduc and E.~Sokatchev,
 Class.\ Quant.\ Grav.  9 (1992) 361.
\bibitem{Delduc:1992fc}
F.~Delduc, E.~Ivanov and E.~Sokatchev,
 Nucl.\ Phys. B384 (1992) 334.
\bibitem{Howe:2000vk}
P.~S.~Howe, A.~Kaya, E.~Sezgin and P.~Sundell,
Nucl.\ Phys. B587 (2000) 481.
\bibitem{dbranes}
M.\ Cederwall, A.\ von Gussich, B.E.W.\ Nilsson and A.\
Westerberg,
Nucl.Phys. B490 (1997) 163.\\
M.\ Aganagic, C.\ Popescu and J.H.\ Schwarz,
Phys.Lett. B393 (1997) 311.\\
M.\ Cederwall, A.\ von Gussich, B.E.W.\ Nilsson, P.\ Sundell and
A.\ Westerberg, Nucl.Phys. B490 (1997) 179.\\
E.\ Bergshoeff and P.K.\ Townsend, Nucl.Phys.  B490 (1997) 145.
\bibitem{hstring}
M. Tonin, Phys. Lett.  B266 (1991) 312;  Int. J. Mod. Phys. A7
1992 6013.
\bibitem{membra}
P.~Pasti, D.~Sorokin and M.~Tonin,
 Nucl.\ Phys. B591 (2000) 109;
``Geometrical aspects of superbrane dynamics'', hep-th/0011020.
\end{thebibliography}
\end{document}